%% file: main.tex
\documentclass[letterpaper, 10 pt, conference]{ieeeconf}  

\IEEEoverridecommandlockouts                              
\usepackage{graphics} 
\usepackage{epsfig} 
\usepackage{amsmath} 
\usepackage{amssymb}  
\usepackage{physics}
\usepackage{qcircuit}
\usepackage{multirow}
\usepackage[hang,small,bf]{caption}    

\usepackage{tikz}	
\usetikzlibrary{backgrounds,fit,decorations.pathreplacing}  

\usepackage{url}
\usepackage[ruled, vlined, linesnumbered]{algorithm2e}
\usepackage{verbatim} 
\usepackage{soul, color}
\usepackage{lmodern}
\usepackage{fancyhdr}
\usepackage[utf8]{inputenc}
\usepackage{fourier} 
\usepackage{array}
\usepackage{makecell}
\usepackage{subcaption}
\usepackage{tabularx}

\SetNlSty{large}{}{:}

\makeatletter

\newcommand{\Rom}[1]{\expandafter\@slowromancap\romannumeral #1@}
\makeatother

\title{\LARGE \bf
Black-Litterman Portfolio Optimization with Noisy Intermediate-Scale Quantum Computers
}

\usepackage{authblk}

\author[1,2,3,*]{Chi-Chun Chen\thanks{* r08222060@ntu.edu.tw}}
\author[4,5]{San-Lin Chung}
\author[1,3,6,$\dagger$]{Hsi-Sheng Goan\thanks{$\dagger$ goan@phys.ntu.edu.tw}}
\affil[1]{Department of Physics, National Taiwan University, Taipei 106319, Taiwan}
\affil[2]{MediaTek, Hsinchu 30078, Taiwan }
\affil[3]{Physics Division, National Center for Theoretical Sciences, Taipei 106319, Taiwan}
\affil[4]{Department of Finance, National Taiwan University, Taipei 106319, Taiwan}
\affil[5]{Department of Digital Financial Technology, Chang Gung University, Taoyuan City 33302, Taiwan}
\affil[6]{Center for Quantum Science and Engineering, National Taiwan University, Taipei 106319, Taiwan}

\begin{document}

\maketitle
\thispagestyle{plain}
\pagestyle{plain}

\begin{abstract}
In this work, we demonstrate a practical application of noisy intermediate-scale quantum (NISQ) algorithms to enhance subroutines in the Black-Litterman (BL) portfolio optimization model. As a proof of concept, we implement a 12-qubit example for selecting 6 assets out of a 12-asset pool. Our approach involves predicting investor views with quantum machine learning (QML) and addressing the subsequent optimization problem using the variational quantum eigensolver (VQE). The solutions obtained from VQE exhibit a high approximation ratio behavior, and consistently outperform several common portfolio models in backtesting over a long period of time. A unique aspect of our VQE scheme is that after the quantum circuit is optimized, only a minimal number of samplings is required to give a high approximation ratio result since the probability distribution should be concentrated on high-quality solutions. We further emphasize the importance of employing only a small number of final samplings in our scheme by comparing the cost with those obtained from an exhaustive search and random sampling. The power of quantum computing can be anticipated when dealing with a larger-size problem due to the linear growth of the required qubit resources with the problem size. This is in contrast to classical computing where the search space grows exponentially with the problem size and would quickly reach the limit of classical computers. 

\end{abstract}


\section{INTRODUCTION}
\begin{center}
\begin{figure*}[h]
    \includegraphics[width=0.95\linewidth]{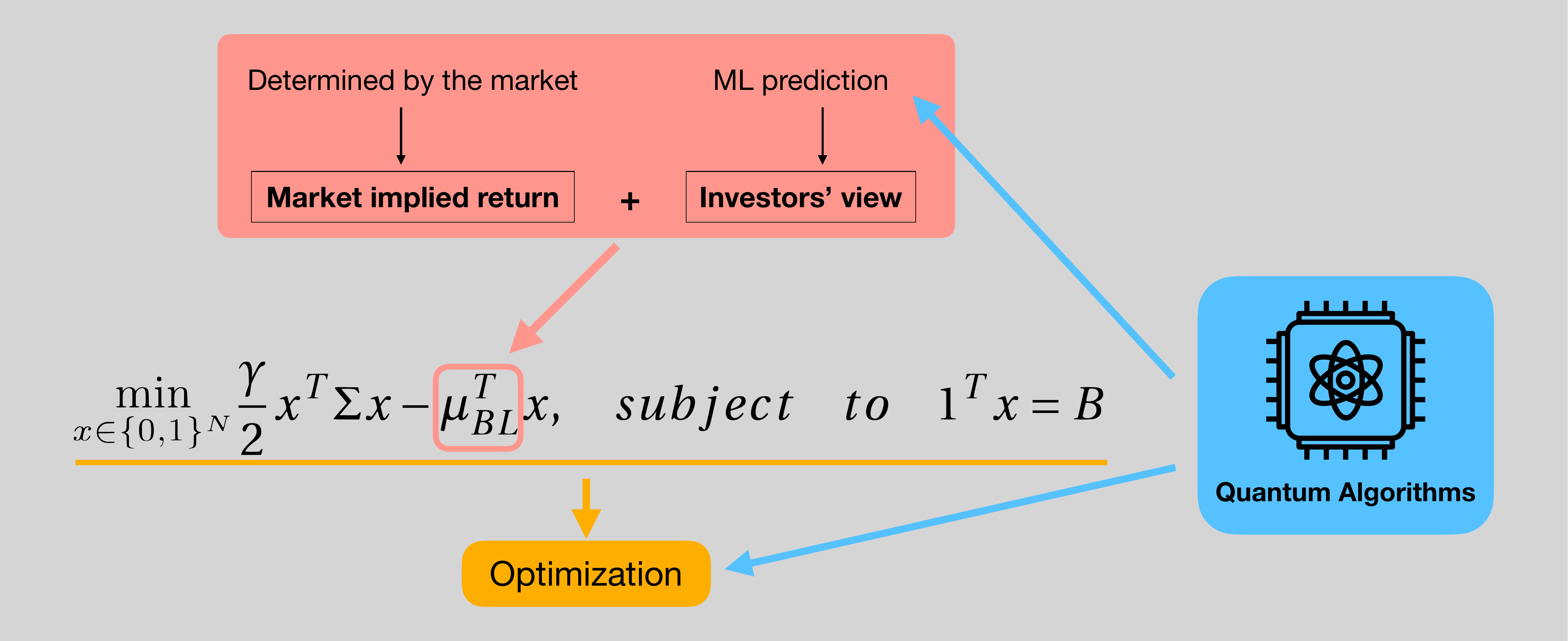}
\caption{Summary of the BL model and how quantum algorithms applied.}
\label{fig:bl_sum}
\end{figure*}
\end{center}

Portfolio optimization is a combinatorial optimization (CO) problem in finance that seeks to identify a set of assets from a pool to maximize returns while minimizing risk\cite{Markowitz1952,perold1984large}. For a continuous capital allocation case, this problem boils down to inversion of a positive semi-definite matrix (the covariance matrix), and is solvable in polynomial time. However, for a discrete case (i.e. the capital is divided into blocks), this problem becomes a combinatorial optimization (CO) problem, whose search space grows exponentially with problem size, and belongs to category of NP-hard in complexity theory. Since efficient classical algorithms are lacking for this type of problems, exploration of quantum computing as a potential solution\cite{orus2019quantum,alcazar2020classical,egger2020quantum} is prompted. Recently, quantum computers with larger scale and better fidelity have came into existence. Several studies have shown potential capability for hybrid quantum-classical methods, which is a strong candidate for application of NISQ computers, to solve portfolio optimization.

The most well known model for portfolio optimization is the modern portfolio theory (MPT) model, introduced in 1952 by Harry Markowitz\cite{Markowitz1952}. MPT is also called the mean-variance model since it quantifies return and risk as mean and (co)variance of daily return, respectively. For a total N assets pool, a portfolio is represented by a vector $x\in\{0,1\}^N$, where $1/0$ denotes buying/not buying an asset. The return of a portfolio can be written as $\mu x$, where $\mu$ is the return vector that represents the expectation value of returns of each asset based on historical mean. On the other hand, risk is defined as the covariance of daily returns, and can be written as $x^T\Sigma x$, where $\Sigma$ is the covariance matrix of the returns. The goal is to maximizes the return while minimizing the risk, and it can be achieved by minimizing the quadratic function
\begin{equation}
\label{qubo}
    \frac{\gamma}{2}x^T\Sigma x-\mu^T x,\quad subject\quad to \quad 1^Tx=B.
\end{equation}
$\gamma$ is the risk aversion coefficient that specifies the importance between minimizing risk and maximizing return. B is the budget constraint that indicates the total number of asset one should buy, and can be implement by adding a penalty term to the objective function, i.e.
 \begin{equation}
\label{qubo_penalty}
    \frac{\gamma}{2}x^T\Sigma x-\mu^T x - \lambda(1^Tx-B)^2.
\end{equation}
The problem is thus a quadratic unconstrained binary optimization (QUBO).

Several studies in the realm of quantum computing have focused on solving discrete portfolio optimization with MPT using VQEs\cite{cerezo2021variational,kandala2017hardware}, and quantum annealing (QA)\cite{alcazar2020classical}. However, despite making significant contributions to the framework of quantifying return and risk for mathematically optimizing the portfolio, MPT has shown limitations in practical cases. The MPT solution has been found to result in issues such as unintuitive and highly-concentrated portfolios, large short positions, and sensitivity to input\cite{black1992global,idzorek2007step}. From a practical point of view, the Black-Litterman model, developed in 1992 at Goldman Sachs by Fischer Black and Robert Litterman\cite{black1992global}, may be more appealing.

The main difference between BL and MPT model is that BL adopt the combined return vector $\mu_{BL}$ as its return term. $\mu_{BL}$ is a statistical combination of two distributions, market implied return and investor's views. Market implied return is obtained through reverse optimization of the market capitalization weight, meaning that the optimal solution is given as the the capitalization weight, and the corresponding returns of each asset that lead to this solution are then calculated. Thus market implied return is entirely determined by the market itself. Investor's view, on the other hand, can be determined in many ways, for example human predictions or other prediction models. Fortunately, previous studies have shown a way to divide these views into categories of trends\cite{meucci2010black} (e.g. very bullish, bullish, bearish, very bearish), which we can leverage classification methods in quantum machine learning (QML). We explore several quantum and classical classification models, eventually obtained $\mu_{BL}$ with the prediction of the quantum kernel method for support vector machine (QSVM)\cite{havlivcek2019supervised} due to the higher accuracy.

After $\mu_{BL}$ is determined, the optimization problem is similar to that of in MPT model\ref{qubo}. The QUBO problem is mapped to Ising Hamiltonian and solved with VQE using different types of ansatz (e.g. heuristic, QAOA). Once the circuits are optimized, we point out that one can—and should—use very few final samplings to find high-quality solution. An intuition of this argument is that we will have to check the eigenvalue of each sampled eigenstate to find the best among them, and thus if the amount of samples is too large, we could just perform same amount of random samplings and will end up with very high probability to find a solution with similar ARs. Moreover, since the search space of this problem grows exponentially, the sample size we are able to check is always a small portion of it for a large size problem.

Since the scale of real quantum device today are not able to solve discrete portfolio optimization problems beyond classical computer limit (and quantum computers cannot be efficiently simulated classically), the main purpose of this work is to provide a procedure of enhancing subroutines in BL model with NISQ algorithms. We demonstrate 12 and 16 qubit case to show the capability of obtaining solutions with good backtesting performance, quantum advantage should be anticipated once we have quantum hardware with size and fidelity beyond classical capacity.

\section{Financial Model and Definition}
\subsection{Black-Litterman model}
BL model can be generally understood as a method to construct a new combined return vector $\mathbf{\mu_{BL}}$, which is a combination of market implied return and investors' view, to replace the $\mu$ in equation \ref{qubo}, and do portfolio optimization. The market implied return $\mathbf{\Pi}$ is the return we should have if the market capitalization weight $\mathbf{w_{mkt}}$ is the optimal solution for a portfolio optimization, and is calculated via reverse optimization:
\begin{equation}
\label{impliedreturn}
    \mathbf{\Pi}=\gamma\mathbf{\Sigma w_{mkt}},
\end{equation}
where $\gamma$ is the risk aversion coefficient as in equation (\ref{qubo}) that depends on investor's preference. For a general case, we choose $\gamma$ based on the ratio of excess return and covariance of the market index over the pass 10 years. Since $\mathbf{\Pi}$ is fully determined by the market, we can incorporate quantum technology into BL model via investor’s views. Investor's views can be quantized by the following three terms : \\
$\mathbf{P}$ :  a matrix that identifies the assets involved in the views.\\
$\mathbf{Q}$ :  the view vector.\\
$\mathbf{\Omega}$ : a diagonal matrix representing the uncertainty of views.\\
For example, if we have a view about two assets A and B, say A will have a return of 0.05 with uncertainty 0.001, and B will have a return of 0.1 with uncertainty 0.0005. The corresponding matrices are written as 
\begin{equation*}
\mathbf{P}=
    \begin{bmatrix}
        1 & 0\\
        0 & 1
    \end{bmatrix}
,\quad \mathbf{Q} = 
    \begin{bmatrix}
        0.05\\
        0.1
    \end{bmatrix}
,\quad \mathbf{\Omega} = 
   \begin{bmatrix}
        0.001 & 0\\
        0 & 0.005
    \end{bmatrix}
\end{equation*}
Note that there are also relative views such as return of A asset will out perform B by some amount, however, in this work we consider only a direct view of every asset (thus $\mathbf{P}$ is an identity matrix). In general, these views can be generate from any kind of resource, including financial analysts, news, or personal experience...etc. However, to take advantage of the quantum classification methods, we should treat the views quantitatively. According to \cite{meucci2010black}, the uncertainty of a view about $k^{th}$ asset can be taken as proportional to the variance of the asset itself. We can thus write
\begin{equation}
\label{omega}
    \omega_k \equiv \tau*\mathbf{p_k\Sigma p_k^T},
\end{equation}
where $\omega_k = \Omega_{kk}$, $\mathbf{p_k}$ is the $k^{th}$ row of $\mathbf{P}$, and $\tau$ is a constant that will not affect the final combined return vector if we use equation \ref{omega} to construct $\mathbf{\Omega}$\cite{idzorek2007step}.
 For the view vector $\mathbf{Q}$, \cite{meucci2010black}also suggested that we can set the $k^{th}$ view in $\mathbf{Q}$ to be
\begin{equation}
\label{viewk}
    q_k \equiv (\mathbf{P\Pi})_k + \eta_k \sqrt{\mathbf{p_k\Sigma p_k^T}},
\end{equation}
where $\eta_k \in \{-2, -1, 1, 2\}$ denotes the view as "very bearish", "bearish", "bullish", and "very bullish" respectively. Here, we further defined $\mathbf{\eta}$ for two binary classification models:
\begin{equation}
\label{Eeta}
    \eta_k=s_1s_2Y_1Y_2, 
\end{equation}
where $Y_1 \in \{+1, -1\}$, $Y_2 \in \{1, 2\}$, and $s_i$ is the accurate rate of the testing data of the classifier for $Y_i$. Now, $\eta_k \in [-2,2]$ since we have multiplied it with a scale $s_1s_2$. This means that we will have a stronger view for the classifier with higher accuracy, and vice versa. After we constructed these matrices, the combined return vector is calculated with the result formula in the original paper \cite{black1992global},
\begin{equation}
\label{muBL}
   \mathbf{\mu_{BL}}=\left[(\tau\mathbf{\Sigma})^{-1}+\mathbf{P^T\Omega^{-1}P}\right]^{-1}\left[(\tau\mathbf{\Sigma})^{-1}\mathbf{\Pi}+\mathbf{P^T\Omega^{-1}Q}\right].
\end{equation}
 By replacing $\mathbf{\mu}$ with $\mathbf{\mu_{BL}}$ in equation (\ref{qubo}), the objective function now becomes 
 \begin{equation}
 \label{quboBL}
     \frac{\gamma}{2}x^T\Sigma x-\mu_{BL}^T x,\quad subject\quad to \quad 1^Tx=B,
 \end{equation}
and the corresponding QUBO form is 
\begin{equation}
 \label{quboBL_penalty}
     \frac{\gamma}{2}x^T\Sigma x-\mu_{BL}^T x + \lambda(1^Tx-B)^2.
 \end{equation}

\subsection{Logarithmic  Return}
Many of the previous quantum finance studies either applied simple return, i.e. $R_i = \frac{p_i-p_{i-1}}{p_{i-1}}$, by default or without giving specific form for return. However, in finance, log return is more commonly used due to it's statistical properties (follows normal distribution) and the ability of capturing compound return (time additive). Log return is defined as 
\begin{equation}
\label{logre}
r_i:=log(\frac{p_i}{p_{i-1}}),
\end{equation}
and when we add them along a time sequence $t=0$ to $t=T$, it becomes $r_{0T}=log(\frac{p_T}{p_0})$. The rate of return (RoR) during this period is $e^{r_{0T}}-1$. On the other hand, the sum of simple return will not capture the compound return. Thus, it is more preferable to adopt log return when we are optimizing the data for a buy and hold strategy.

\subsection{Effective $\gamma$}
The risk aversion coefficient $\gamma$ is another thing that is also not discussed in quantum finance studies. In the discrete optimization case, the budget constraint, $1^Tx=B$, is needed otherwise the solutions are not comparable. However, when we construct a portfolio in real cases, $x$ should always be re-normalized to 1 after we obtain the solution, i.e. $x_r=\frac{1}{B}x$. Since the variance in the objective function (equation (\ref{quboBL})) is quadratic and the return is linear, the re-normalization will result in equivalently solving for a risk aversion coefficient $\frac{\gamma}{B}$. For cases that $\gamma$s are chosen randomly, this may not matter, nevertheless, if $\gamma$ is estimated with historical data and has financial meaning, which is the case here, it should be carefully considered. We thus propose an effective $\gamma$:
\begin{equation}
\label{effgamma}
   \gamma_{eff}=\frac{\gamma}{B},
\end{equation}
where $\gamma$ can be estimated with, for example, the ratio of the risk premium to the variance of index over the pass 10 years, i.e. 
\begin{equation}
\label{effgamma}
   \gamma =\frac{\mu-r_f}{\sigma^2},
\end{equation}
where $\mu$ and $\sigma^2$ are the mean return and variance of the index, and $r_f$ is the mean risk free rate. Here we refer $r_f$ to the 13 week treasury bill (Yahoo Finance : $\hat{ }IRX$).
We use $\gamma_{eff}$ rather than $\gamma$ for solving equation (\ref{quboBL}), and when we re-normalize the optimal solution such that the sum of portfolio weights is 1, it is an optimal solution for the risk aversion coefficient $\gamma$ that we want.

\subsection{Index}
In the framework of BL model, the market cap weighted index is assumed to be efficient at present. For a problem of asset pool that are large enough and are all included in $S\&P500$, we can refer the index to the $S\&P500$ index. In our case, however, we have to construct the corresponding market weighted index for our assets pool. Here, 12 individual stocks are considered, we determined the market cap by the data at closest time to the end time of the training period. Since the market cap weight is in unit of money, we first assume a total 100 unit of money (i.e. the initially normalized index level), and calculate the shares for each asset at the time of market cap data. The shares then holds as a constant, and price of the index at different time is determined by 
\begin{equation}
\label{index}    
    P^t_{index}=\sum_i P^t_iS_i,
\end{equation}
where $P_i$ is the price of $i$ asset, and $S_i$ is the shares of $i$ asset.

\subsection{Walk-Forward Backtest}
The walk-forward backtest is a common backtest method which one continuously move a window, which is consist of a training period and a following testing period, along a long period of time.
We performed a walk-forward backtest for total time period from 2008/01/01 to 2021/12/31, with a 260 week (about 5 year) training period and 52 week (about 1 year) testing period, and a moving interval of 52 week to analyze the performance of BL model. Note that our moving interval is identical to the testing period, thus this backtest can also be seen as a dynamic portfolio optimization without considering the transaction fee.

\section{Method and Algorithms}
In this section, we discussed the method and quantum algorithms we applied and the circuits we used for BL model, including A. quantum classifiers for investors' views, and B. penalty tuning for implementing constraint, C. VQEs for optimization, and D. data sources and preprocessing.

\subsection{Quantum classifiers}
The goal for quantum classifiers is to classify $Y_1\in\{-1, 1\}$ and $Y_2\in \{1, 2\}$ in equation (\ref{Eeta}) after $t$ trading units (e.g. days/weeks) in the future for each asset in our pool, and predict the given feature vector $\mathbf{X}$ at the latest date of training period to construct $\eta s$ via equation (\ref{Eeta}), and thus the view matrix $\mathbf{Q}$. We can represent the feature of any given date $D$ with financial indicators of that day, and labeled $Y1$ and $Y2$ according to the mean (log) return $\overline{r}$ of the $t$ trading units since $D$. The labels are given by the following rules:
\begin{equation}
\label{labelY}
\begin{aligned}
    Y_1 & = &
    \left\{ 
        \begin{array}{cc}
         -1,&  if \quad \overline{r} <0 \\
         +1,&  if \quad \overline{r} \ge 0, 
        \end{array}
    \right. \\
    Y_2 & = &
    \left\{ 
        \begin{array}{cc}
         \ \ 1,&  if \quad \frac{|\overline{r}|}{\sigma'} < 1 \\
         \ \ 2,&  if \quad \frac{|\overline{r}|}{\sigma'} \ge 1, 
        \end{array}
    \right.
\end{aligned}
\end{equation}
where $\mathbf{\sigma'}=\frac{\sigma}{\sqrt{t}}$, and $\sigma$ is the standard deviation of $\{r\}$ of the corresponding assets. Since the testing period is 52 weeks, $t=52$ here.

\begin{center}
\begin{figure}[h]
    \begin{subfigure}{0.48\textwidth}
    \includegraphics[width=\linewidth]{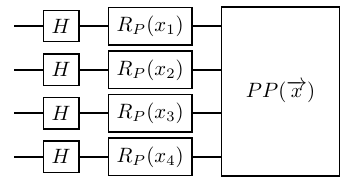}
    \caption{} \label{fig:P_PP}
    \end{subfigure}%
    
    \begin{subfigure}{0.48\textwidth}
    \includegraphics[width=\linewidth]{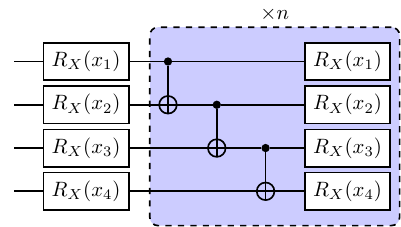}
    \caption{} \label{fig:simple}
    \end{subfigure}%
    
\caption{(a) The Pauli feature map for data encoding, where $P=X,Y,Z$ is the rotation Pauli gate. (b) The simple embedding feature map for data encoding.}
\label{fig:featuremap}
\end{figure}
\end{center}

In this work, we implemented two kinds of common quantum supervised learning models, which are quantum neural network (QNN)\cite{mitarai2018quantum,farhi2018classification}, and quantum kernel method for support vector machine (QSVM)\cite{havlivcek2019supervised}, and compared them with classical NN and SVM. Both quantum classifiers rely on a quantum circuit feature map $\mathcal{U}_{\Phi}(\overrightarrow{x})$ that encodes classical data into Hilbert space, but the training part works in different ways. For QNN, the feature map circuit is directly followed by an ansatz circuit with trainable parameters. Every data point that are encoded with the feature map and go through the unitary transformation of the ansatz, can be measured in computational basis and output a collection samples $\{b\}$. We interpret these measurement bit strings with a parity function that maps the samples to $sum(b)\ mod\ 2$, resulting in a binary probability distribution. We then use classical optimizer to minimize the squared loss function and update the parameters in the ansatz circuit. For QSVM, the quantum kernel can be represented by $\mathcal{U}_{\Phi}(\overrightarrow{x})^\dagger \mathcal{U}_{\Phi}(\overrightarrow{x})$, which the same feature maps as shown in Fig \ref{fig:featuremap}, followed by its own dagger. We measure the probability of $\ket{0}^{\otimes n}$, which has been shown in \cite{havlivcek2019supervised} to be equivalent to the inner product of the data points in feature space. Thus, we can apply this quantum kernel to the classical support vector machine (SVM).

We search among several kinds of feature map circuits, including the simple embedding \ref{fig:simple} and the Pauli feature map \ref{fig:P_PP}, with different Pauli gates (e.g. [Z, YY], [Y, ZZ]...etc.) and entanglement structure (e.g. linear, circular, and full), finding that in our case, a simple embedding feature map without any entanglement structure (n=0) is enough to obtain high testing accurcies. QSVM with simple embedding feature map turns out to be slightly better than classical SVM with Gaussian kernel. The Pauli feature map, on the other hand, includes two-qubits rotational gates. It is more complex but did not work well on this task since we are training with only a small size and simple feature data. For predictions of more complex situations, which the feature dimension may be larger and more complex, one may be able to leverage the complexity of entangled quantum feature map. On the other hand, QNN and NN did not perform as well as SVM based ML model here.

\begin{center}
\begin{figure}[h]
    \includegraphics[width=0.95\linewidth]{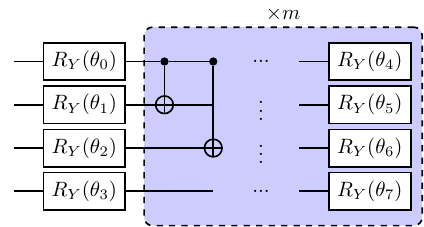}
\caption{The parameterized ansatz for QNN, where $\theta s$ are trainable parameters, and the omitted part are fully connected CNOT gates. We choose m=2 for QNN here.}
\label{fig:ansatz}
\end{figure}
\end{center}

\subsection{Penalty tuning}
\begin{center}
\begin{figure}[h]
    \includegraphics[width=0.95\linewidth]{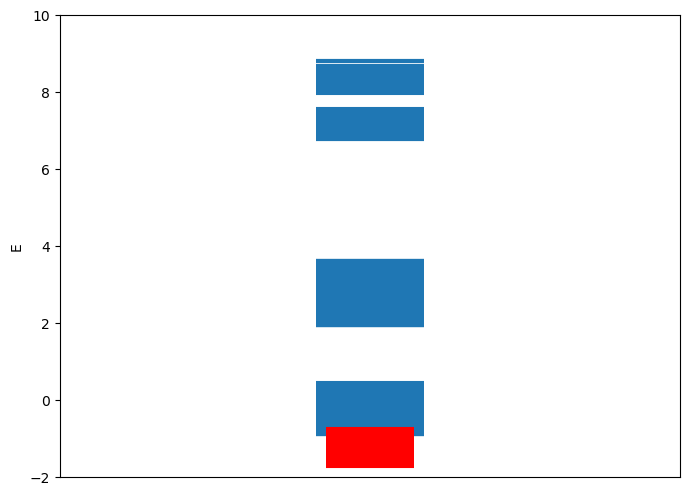}
\caption{Energy eigenvalue distribution of Hamiltonian : $\frac{\gamma}{2}x^T\Sigma x-\mu_{BL}^T x + \lambda(1^Tx-B)^2$, with $\lambda=1$. The red eigenstates are ones that obeys the constraint, and blue eigenstates are ones don't. The lowest blue state should be at least higher than half of the red states to successfully implement the constraint. Eigenvalues above 10 are omitted.}
\label{fig:penalty}
\end{figure}
\end{center}

Before actually solving the QUBO objective function, i.e. equation (\ref{quboBL_penalty}), we must first determine a proper penalty $\lambda$. $\lambda$ should be large enough to implement the constraint, but not unnecessarily large to dominate the Hamiltonian, making the risk-return part requires extreme precision to solve. Moreover, a large $\lambda$ also directly increases the coefficient of the Hamiltonian, and thus we need more measurements (circuit repetitions) to evaluate the expectation value to same precision. Here we choose $\lambda=1.0$ by empirically analyze the distribution of eigenvalue of the Hamiltonian. We believe that choosing $\lambda$ such that the eigenvalue of the lowest eigenstate that doesn't satisfy the constraint is at least larger than the $50\%$ but less than $100\%$ of the eigenvalue of eigenstates that satisfy the constraint, as shown in Fig \ref{fig:penalty}, will be preferable for VQEs.

\subsection{VQEs}
\begin{center}
\begin{figure}[h]
    \includegraphics[width=0.95\linewidth]{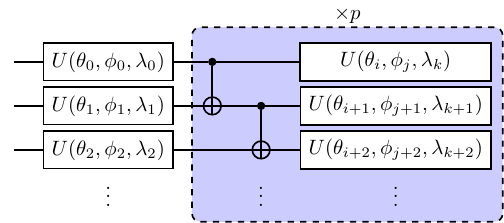}
\caption{The parameterized ansatz for VQE, where $U$ is a general rotation gate represented by ($RY(\theta)$, $RZ(\phi)$, $RY(\lambda))$, and $\theta s, \phi s, \lambda s$ are trainable parameters. The circuit repetitions are chosen $p=4$ in our case, with each layer (the colored part) being linearly entangled.}
\label{fig:vqe}
\end{figure}
\end{center}

 We solved equation (\ref{quboBL_penalty}) with gate-based quantum computer simulator using a heuristic circuit VQE, and QAOA\cite{farhi2014quantum}. QAOA is also a type of VQE that has a specific variational form that depends on the problem Hamiltonian, and is related to the quantum adiabatic algorithm. QUBO problems are natively solvable for quantum computers since the objective function can easily be mapped to an Ising Hamiltonian by $x=\frac{1-Z}{2}$, where $x\in\{0,1\}$ and $Z\in\{1,-1\}$ is the Pauli Z operator. The mapped Ising Hamiltonian of equation (\ref{quboBL_penalty}) is 
\begin{equation}
\label{hamiltonian}
    H = \sum_{i}\sum_{j>i}h_{ij}Z_iZ_j + \sum_{i}h_iZ_i + C,
\end{equation}
where $h_{ij}$ and $h_i$ are coefficients of corresponding terms and $C$ is the offset constant. This Ising Hamiltonian should in general be fully connected, unless the covariance between some of the assets are zero. With this problem Hamiltonian, we construct the heuristic ansatz for VQE shown in Fig \ref{fig:vqe}, and QAOA ansatz is built following the instruction stated in \cite{farhi2014quantum}. From there, we then update the parameters in the circuit with Sequential Least Squares Programming (SLSQP) optimizer in order to minimize the Hamiltonian expectation value. After convergence, the optimal parameters is then fixed in the ansatz to perform final samplings, and the solution is chosen as the outcome bit string with lowest eigenvalue among the final measurements. We take only 5 measurement shots as final samplings here. There are two reasons to use such few shots, first is that once we optimized the parameter in the ansatz successfully, the probability distribution should concentrate on low energy eigenstates, thus we should expect needing only few final samplings to get a good solution with high probability. Secondly, this is also to mimic the situation of solving a large size problem, which the final shots we can get is always a tiny portion of the total search space, and should always not exceed the search space (otherwise one should rather perform a exhaustive search). Since the search space grows exponentially, a large amount of final samplings (compare to search space) is neither justified nor possible. We further analyzed the probability of getting a "good" solution with random sampling in appendix A. We show that the probability of reaching such AR within 5 final samplings is very low, and provide a upper bound for maximum number of final sampling. 

The heuristic VQE form we used requires $3N(p+1)$ parameters for an N assets optimization problem, where p is the repetition of the circuit. For QAOA, the entanglement has to be fully connected since the problem itself is. It might be more hardware demanding for most cases, but requires only $2p$ parameters, and the performance is guaranteed to improve (or at least the same) as $p$ increases\cite{farhi2014quantum}. We choose $p=4$ for heuristic VQE and $p=8$ for QAOA, and take 10 and 500 initial parameter guesses for each respectively. The initial guess that is eventually optimized to the lowest expectation value will served as the optimal parameter and used to perform final sampling.



\subsection{Data}
We applied quantum algorithms to BL model and implement it for real world stock data, with walk-forward back test framework, i.e. a 262-week (5 years) in-sample period and a 52-week (1 year) out-sample period, starting from $2008-01-01$ and ends at $2021-12-31$, creating a total 9 continuous time segments. The training and testing stock price data are all downloaded from Yahoo finance weekly data, and the market capitalization data is obtained from \cite{compan}. The financial indicators for predicting investor's views (TABLE \ref{table:indicators}) are downloaded from "Federal Reserve Bank of St. Louis" \cite{fred1997}. The asset pool contains 12 individual stocks randomly chosen from $S\&P500$ companies of different industries. 

It requires several preprocessing before the financial indicators can become a feature vector for prediction tasks. \\
1. Backward moving average with a 3-week rolling window.\\
2. Standard normalization.\\
3. PCA reduce to 4 dimension.\\
4. Re-normalize each feature to $(0,2\pi]$.\\
We then label the these features of each week according to the rule in equation (\ref{labelY}). In order to improve the "trainability" of our data, we remove both feature vectors if they are too close in feature space but having different labels. The main idea of doing this is that we don't want those feature vectors that give no information, or more specifically, require extremely high precision to distinguish each other, to distract our models. The feature vectors can then be encoded into the quantum circuit via our quantum feature map to train predictive models. We further extend the procedure to a global portfolio optimization case in Appendix B, which the asset pool contains 16 ETFs that approximately represents the global economy.

\input{table/indicators}


\section{RESULTS \& DISCUSSION}
\subsection{Investor's views}
\begin{center}
\begin{figure}[h]
    \includegraphics[width=0.95\linewidth]{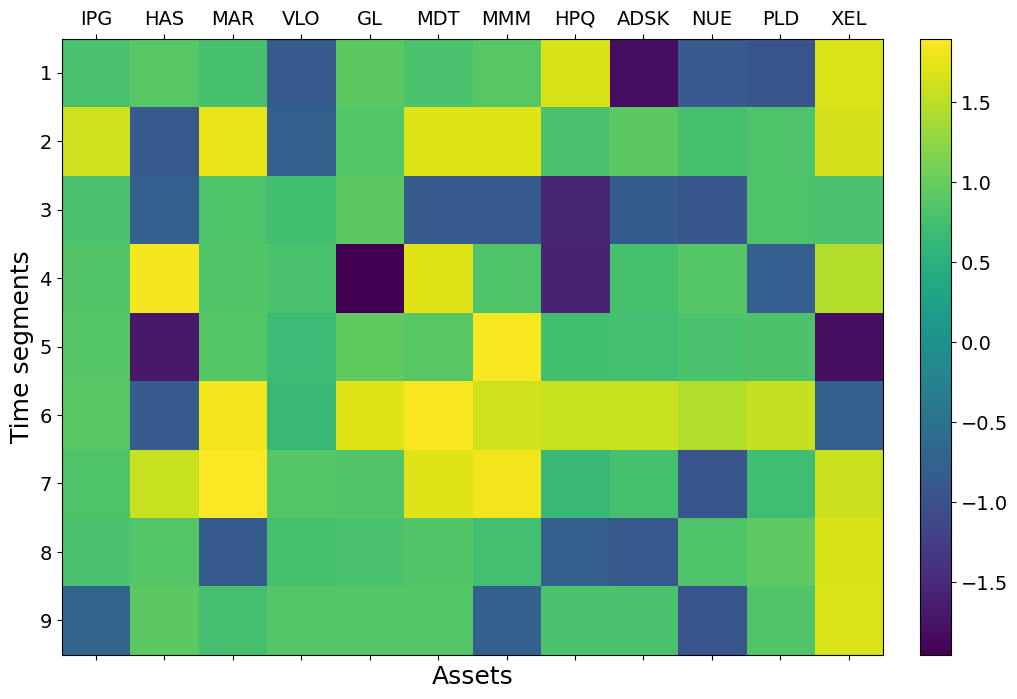}
\caption{$\eta$ predictions of QSVM.}
\label{fig:etas}
\end{figure}
\end{center}

We here showcase in TABLE \ref{table:qml_score1} the average prediction accuracy for $Y_1$, and $Y_2$ over total 9 time segments of each asset. We found that in average, $QSVM\approx SVM > NN > QNN$ in terms of testing accuracy. QSVM is also much faster to train compare to QNN. Here, we chose QSVM models to predict the investor's view for future optimizations. The corresponding $\eta s$ predicted by QSVM for each assets is shown as a colored matrix in Fig \ref{fig:etas}.

\input{table/qml_score1.tex}
Since the problem size here is not large, both classical and quantum kernel performed well. However, we may look forward to potential advantage for quantum kernel in predictions for more complex situations, as in real world use cases.

\subsection{Optimization}
As stated above, we applied (a) VQE circuit in Fig \ref{fig:vqe} with $p=4$, (b) QAOA with $p=8$, and (c) exhaustive search to solve all QUBO problems. Before we show the results, we have to first re-define the standard of measuring how good a solution is. In general, people usually refer to the approximation ratio(AR), i.e. $\frac{E}{E^*}$ for a minimization tasks, or the relative error, i.e. $\frac{|E-E^*|}{E^*}$, where $E$ is the eigenvalue obtained from the algorithm, and $E^*$ is the exact minimum eigenvalue. However, for a portfolio optimization problem, $E^*$ may often be around $0$, and thus results in unreasonably large ARs or relative errors. Moreover, after we include the $1^Tx=B$ constraint into the penalty term (i.e. equation (\ref{quboBL_penalty})), the energy level of the Hamiltonian splits into chuncks, due to the large amount of extra energy a solution will gain if it violates the constraint (to different degrees), see Fig \ref{fig:penalty} as an example. Thus, if we look at the whole energy range of the Hamiltonian with penalty, which may be orders larger than the one without penalty, and construct an AR based on how close it is to the ground state, the solutions that simply meets the constrain will turn out to have very high AR and thus becomes hard to distinct a good solution from a bad one if they both meet the constraint. We thus here define AR as
\begin{equation}
    AR = \frac{E-E_w}{E^*-E_w} \leq 1,
\label{approx}
\end{equation}
which is equivalent to shifting the worse solution that meets the constraint $E_w$ to $0$. Note that $E^*\leq E \leq E_w=0$, and the solutions that violates the constraint will have negative AR (but positive AR do not necessarily meets the constraint).

\begin{center}
\begin{figure}[h]
    \begin{subfigure}{0.48\textwidth}
    \includegraphics[width=\linewidth]{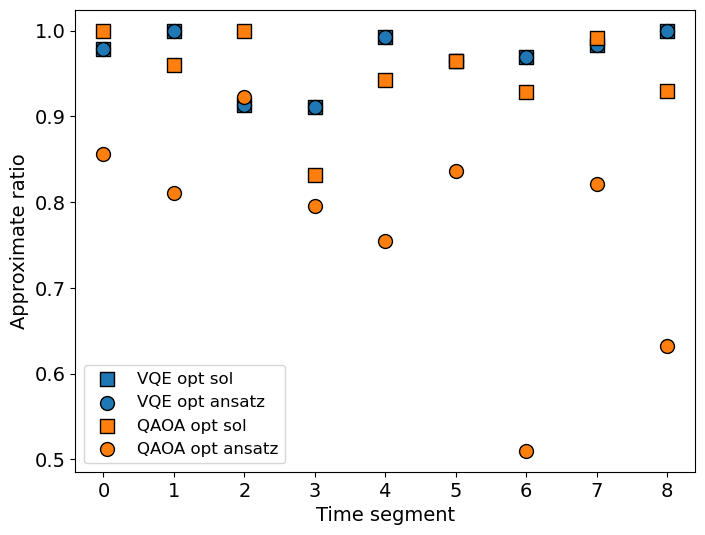}
    \caption{} \label{fig:AR1}
    \end{subfigure}%
    
    \begin{subfigure}{0.48\textwidth}
    \includegraphics[width=\linewidth]{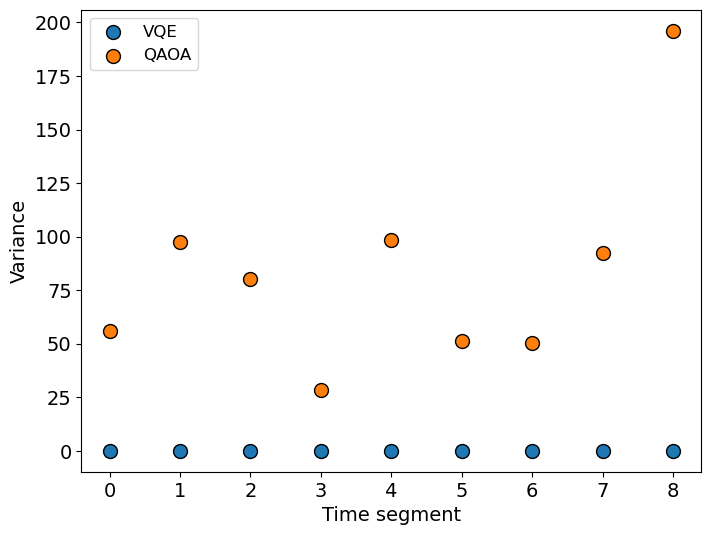}
    \caption{} \label{fig:var1}
    \end{subfigure}%
    
\caption{(a) The AR of optimal ansztzs of VQE/QAOA and the optimal solution obtained by 5 final sampling the corresponding optimal circuits. (b) The variance of the optimal ansatz of VQE/QAOA, calculated by 500,000 sampling.}
\label{fig:optresults}
\end{figure}
\end{center}

The optimization result is shown in Fig \ref{fig:optresults}.
First, we look at the AR of the optimal circuits (\ref{fig:AR1}), which is the corresponding AR of $\expval{H}{\psi(\overrightarrow{\theta^*})}$, where $\psi(\overrightarrow{\theta^*})$ is the ansatz with optimal parameters $\theta^*$. We see that VQE heuristic ansatzs are all solved to at least $0.9$, and the mean of AR is $0.96$. When we perform the final samplings, we found the outcome contains only a single eigenstates, meaning that the optimal ansatz has probability distribution highly concentrated around one or a few eigenstates. On the other hand, the optimal QAOA ansatzs wasn't solve to AR as good as heuristic ansatzs, and the final samplings has a distribution over several eigenstates. However, we are still able to get a relatively good solution from only 5 samplings in our case, but it is not guaranteed. We further analyzed the optimal ansatzs by sampling 500000 shots and calculate the variance of eigenvalues in \ref{fig:var1}. As expected, the variance of VQE heuristic ansatzs are almost 0 and QAOA ansatz is large. Furthermore, although the heuristic ansatz still suffers from local minimum and is not easy to solve it to the ground state, a few number of initial guess is sufficient to solve it to a good AR. For QAOA, the result of optimization varies tremendously with the initial point, and we may usually get bad solutions (AR of ansatz $<0$) if we don't provide enough number of initial guesses. We are not able to determine which ansatz form is better by only looking at the AR and variance of the optimal ansatzs, since they are both able to obtain a solution with good AR within 5 final samplings. However, in terms of the amount of initial guesses, circuit depth and connectivity, and the concentration of probability, we came to a conclusion that VQE heuristic ansatz should be preferred. 


\begin{center}
\begin{figure*}[h]
    \includegraphics[width=0.98\linewidth]{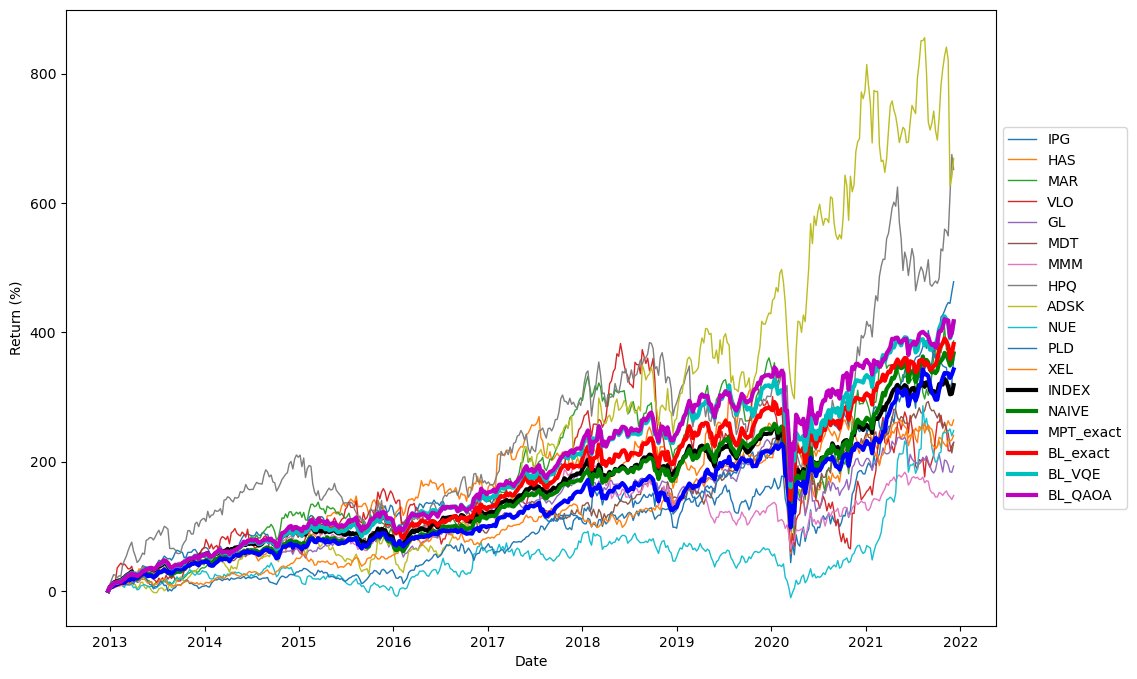}

\caption{Backtesting of market-cap-weighted index (0.107), naive portfolio (0.111),  MPT model exact solution (0.103), BL model exact solution (0.112), BL model VQE solution (0.121), and BL model QAOA solution (0.119), with QSVM view model, for 9 continuous 52-week time segments. The numbers in parentheses represent the mean CERs. The plot considers the capital growth over each time segment for better visualization, but each back test segment could be performed independently.}
\label{fig:backtest_QSVC}
\end{figure*}
\end{center}
\subsection{Backtesting performance}
For backtesting, we adopt the view predicted by QSVM, and the show both the exact solution and the optimal solution obtain by VQE and QAOA. Here, the portfolio performance is measured by certainty-equivalent return (CER):
\begin{equation}
 CER  =  \mu - \frac{\gamma}{2}\sigma^2,
\label{cer} 
\end{equation}
where, $\mu$ is the sample mean of (log) return, and $\sigma$ is the standard deviation of the (log) return. The backtesting period covers 9 continuous 52-week time segments, and is compared to 3 kinds of common portfolios, MPT model, naive portfolio ($1/N$ portfolio), and the market-cap-weighted index. In our case, BL model out performs MPT model in terms of both pure return and mean CER in a long continuous becktesting period. We also show that for VQE/QAOA solutions with high enough ARs could performed similarly as the optimal solution, and even may have a chance to outperform exact solutions. Note that we are not claiming a slightly worse solution will outperform exact solution, but a solution with high AR is enough to perform as good as exact solution. For example in the extended case in Appendix B, the performance is proportional to ARs. This means that we may not necessarily have to solve this problem to the ground state, and we can save computational cost (in terms of Hamiltonian evaluation and classical optimization) by requiring less precision.

\section{CONCLUSIONS}
In this paper, we showcase a practical application of NISQ algorithms to enhance subroutines in the Black-Litterman model. Our results demonstrate better performance compared to several common models over an extended backtesting period. Additionally, the approximated solutions obtained from VQE/QAOA perform similarly to the exact solutions obtained by exhaustive search.

Regarding quantum machine learning, we construct investor views using QNN and QSVM. QSVM, with better testing accuracy and shorter training time, stands out. We anticipate potential advantages for quantum kernels due to their ability to extract complex features in exponential Hilbert space with linearly scaling qubits. This could become valuable for predictions in more complex situations in the future.

In the optimization part, we conclude that the heuristic circuit is preferable over QAOA due to its shallower circuit depth, requiring fewer initial guesses, and exhibiting greater stability in optimization (tendency to concentrate on one or a few eigenstates). Furthermore, we point out that one can—and should—use such few final samplings to find a high-quality solution once the ansatz is optimized. Both variational forms, when optimized, yield solutions with high ARs with only 5 final measurement shots for a 12-qubit problem size. This minimum measurement scheme reduces the number of circuit repetitions in VQE.

Potential quantum advantage is envisioned for larger problems in the future due to the ability of quantum computers to address portfolio optimization problems with linearly scaling qubits. This is in stark contrast to classical computers, which are not able to deal with exponentially growing search spaces.

\bibliographystyle{ieeetr}
\bibliography{ref}


\onecolumn
\appendices
\section{Final samplings}
\label{appendix:final_sample}
Here we estimate an upper bound of final measurements one could use to sample the optimal ansatz. Final measurements greater than $K$ is considered inefficient, and one would rather perform random samplings. Assuming that the probability of randomly sampling one solution with $AR \geq g$ is $P_g$. The probability of getting at least one solution with $AR \geq g$ within $K$ random samplings is 
\begin{equation}
   P^{K}_{g}=1-(1-P_g)^K.
\end{equation}
Thus, the reasonable final measurements one could perform in order to solve the problem to $AR \geq g$ is $K$, such that 
\begin{equation}
   P^{K}_{g} \leq \frac{2}{3},
\end{equation}
according to the standard of bounded-error probabilistic polynomial time (BPP) class algorithms.
In our case, the optimal solution of VQE and QAOA with 5 final samplings is about $0.95$ averaging over 9 time segments. We thus assume $g=0.95$, and $P_g=1.71\times 10^{-3}$ is also a statistical mean over 9 cases. The maximum $K$ for justified final sampling is thus $643$. Moreover, if we conduct only 5 final samplings, the probablity to reach such AR is $P^{5}_{0.95}=0.085$. This means that our VQE/QAOA ansatzs are well solved since we only need 5 final measurement to achieve such AR.

\section{Global portfolio optimization}
\label{appendix:case2}

\begin{figure*}[h!]
\centering
\includegraphics[width=0.98\linewidth]{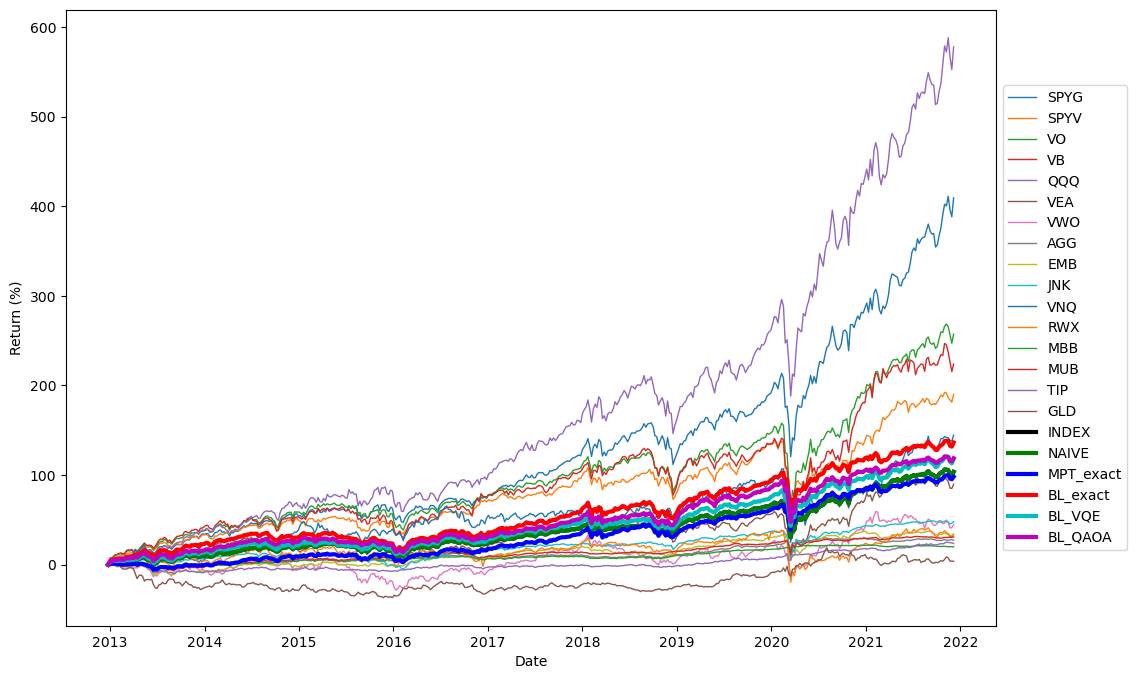}
\caption{Backtesting of a 16 assets portfolio optimization. Including the market-cap-weighted index=naive (0.04), MPT model exact solution (0.048), BL model exact solution (0.051), BL model VQE solution (0.046), and BL model QAOA solution (0.04), with QSVM view model, for 9 continuous 52-week time segments. The numbers in parentheses represent the mean CERs. The plot considers the capital growth over each time segment for better visualization, but each back test segment could be performed independently.}
\label{fig:case2_backtest_QSVC}
\end{figure*}

We perform our BL optimization on another case containing a 16 assets pool that approximately represents the global economy, with budget constraint $B=8$. In this case, the market cap is hard to estimate, we thus consider them equal, the index here is thus identical to a naive portfolio. We applied the QSVC model also with simple embedding feature map with $n=0$ as investor views. The QUBO is then optimized with VQE circuit shown in Fig \ref{fig:vqe} with $p=6$ and 20 random initial guesses, and QAOA with $p=10$ and 500 initial guesses. The solution is the best among 10 final measurements taken from the optimized ansatz. We here show the backtesting results as well as mean CERs.

\section{Real device}
\label{appendix:realdevice}
We execute our optimized VQE heuristic circuits (12 qubits) on IBM Auckland, a 27-qubit superconducting gate based quantum computer, for the purpose of final sampling. The result is shown in Fig\ref{fig:realdevice_VQE} below. We are able to obtain the same solution as simulation result (red circles) within only 5 shots for 6 out of 9 cases, implying that the circuit depth remain suitable for NISQ computers today. On the other hand, the optimized QAOA circuits are much deeper and still have a distribution over a wide range of eigenstates, thus we are not able to gain useful conclusion from performing final measurements on real device.
\begin{figure*}[h!]
\centering
\includegraphics[width=0.85\linewidth]{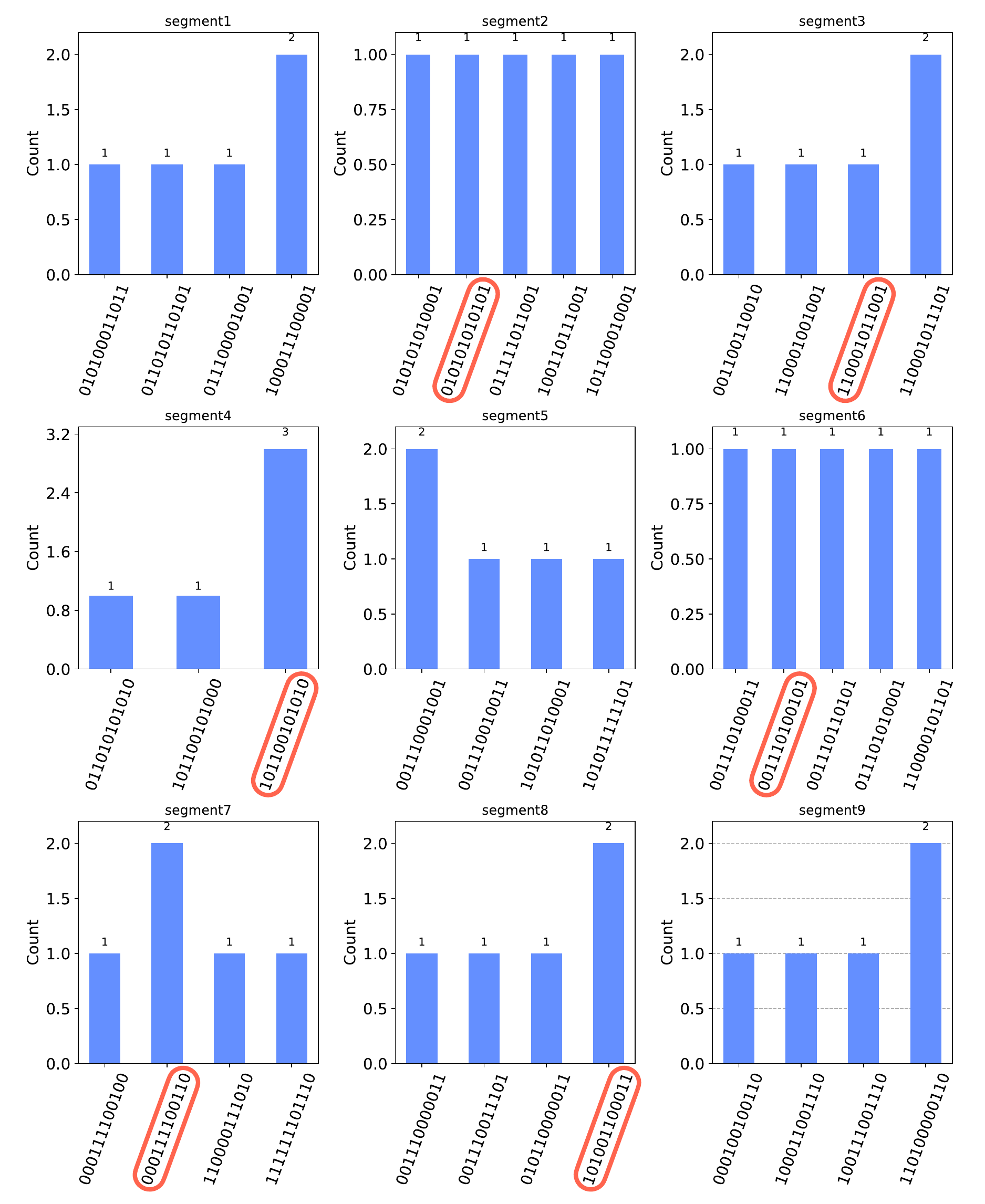}
\caption{}
\label{fig:realdevice_VQE}
\end{figure*}







\end{document}

%% file: table/indicators.tex
\begin{table}
\begin{center}
\begin{tabularx}{0.47\textwidth} { 
  | >{\centering\arraybackslash}X 
   >{\centering\arraybackslash}X
   >{\centering\arraybackslash}X
  | >{\centering\arraybackslash}X | } 
  \hline
  DOW & WILL5000INDFC & VIXCLS\\

  T10Y2Y & T10YIE & DCOILBRENTEU\\
  
  DEXCHUS & DFF & EXPTOTUS \\

  IGREA &  &\\
  \hline
\end{tabularx}
\caption{Financial indicators used as feature for predicting investor's views. Downloaded form Federal Reserve Bank of St. Louis \cite{fred1997}.}
\label{table:indicators}
\end{center}
\end{table}

%% file: table/qml_score1.tex
\begin{table*}[t]
\centering
\resizebox{\textwidth}{!}{
\begin{tabular}{ |c|c c| c c| c c| c c| c c| c c| c c| c c| c c| c c| c c| c c| c c| } 
    \hline
    & \multicolumn{2}{c|}{IPG} & \multicolumn{2}{c|}{HAS}  & \multicolumn{2}{c|}{MAR} & \multicolumn{2}{c|}{VLO} & \multicolumn{2}{c|}{GL} & \multicolumn{2}{c|}{MDT} & \multicolumn{2}{c|}{MMM} & \multicolumn{2}{c|}{HPQ} & \multicolumn{2}{c|}{ADSK} & \multicolumn{2}{c|}{NUE} &  \multicolumn{2}{c|}{PLD} & \multicolumn{2}{c|}{XEL} & \multicolumn{2}{c|}{mean} \\
    \hline
    & $Y_1$ & $Y_2$& $Y_1$ & $Y_2$& $Y_1$ & $Y_2$& $Y_1$ & $Y_2$& $Y_1$ & $Y_2$& $Y_1$ & $Y_2$& $Y_1$ & $Y_2$& $Y_1$ & $Y_2$& $Y_1$ & $Y_2$& $Y_1$ & $Y_2$& $Y_1$ & $Y_2$& $Y_1$ & $Y_2$& $Y_1$ & $Y_2$ \\
   
    \hline
QSVM & 0.92 & 0.9 & 0.98 & 0.88 & 0.97 & 0.88 & 0.92 & 0.84 & 0.96 & 0.93 & 0.92 & 0.94 & 0.93 & 0.91 & 0.9 & 0.85 & 0.9 & 0.91 & 0.89 & 0.95 & 0.91 & 0.92 & 0.98 & 0.83 & 0.93  &  0.9 \\
\hline
QNN & 0.76 & 0.75 & 0.83 & 0.72 & 0.78 & 0.75 & 0.75 & 0.75 & 0.79 & 0.76 & 0.74 & 0.78 & 0.82 & 0.78 & 0.77 & 0.72 & 0.78 & 0.77 & 0.73 & 0.72 & 0.77 & 0.78 & 0.82 & 0.65 & 0.78  &  0.74 \\
\hline
SVM & 0.92 & 0.91 & 0.97 & 0.88 & 0.95 & 0.91 & 0.93 & 0.83 & 0.94 & 0.93 & 0.9 & 0.92 & 0.92 & 0.9 & 0.93 & 0.81 & 0.9 & 0.9 & 0.85 & 0.92 & 0.93 & 0.88 & 0.98 & 0.83 & 0.93  &  0.89 \\
\hline
NN & 0.86 & 0.87 & 0.93 & 0.76 & 0.95 & 0.82 & 0.86 & 0.75 & 0.96 & 0.86 & 0.86 & 0.87 & 0.93 & 0.86 & 0.91 & 0.74 & 0.86 & 0.86 & 0.85 & 0.93 & 0.91 & 0.83 & 0.98 & 0.71 & 0.91  &  0.82 \\
\hline
  
\end{tabular}}
\caption{Testing accuracy of 4 ML models for each assets. Each score is the average of all 9 time segments.}
\label{table:qml_score1}
\end{table*}